\documentclass[b5paper,twoside]{jpconf}  
\usepackage{graphicx}
\usepackage{picinpar}

\begin{document}
\title[PSR-SNR associations]{78 pairs of possible PSR-SNR associations}
\author[Hongquan SU, Qing-Kang Li, Hui ZHU, Wen-Wu TIAN]{Hongquan SU$^{1,2}$, Qing-Kang Li$^1$, Hui ZHU$^2$, Wen-Wu TIAN$^{2}$}
\address{$^1$Department of Astronomy, Beijing Normal University, Beijing 100008, China E-mail: hq\_su@mail.bnu.edu.cn}
\address{$^2$National Astronomical Observatories, Chinese Academy of Sciences, Beijing, 100012, China}

\begin{abstract}
We discuss the criteria to associate PSRs with SNRs, and summary 78 pairs of possible PSR-SNR associations which is the most complete sample so far. We refine them into three categories according to degree of reliability. Statistic study on PSR-SNR associations helps us understand massive star evolution and constrain pulsar's theory models.
\end{abstract}
\section{Criteria for Association}
\subsection{Position}
The associated PSR/SNR should roughly be in the same direction and at the same distance, i.e. the same position. SNRs have  extended structures, so associated PSRs are usually inside SNRs. To determine distances of both PSRs and SNRs are challenge jobs. Sometimes a pair of association listed in our table may have different distance values. The difference might be caused by distance estimate methods and could be reduced by future measurement.
\subsection{PSR¡¯s age}
The measured ages of both the PSRs and SNRs should be the same within the error ranges. The characteristic age of a PSR can be easily got. We compare the real age and the characteristic age of three well-known PSRs. The large age discrepancy between real and characteristic value of three PSRs suggests that pulsar characteristic age is poor age estimator for young pulsars.
\subsection{PWN}
Pulsars steadily dissipate their rotational energy via relativistic winds. Confinement of these outflows generates luminous pulsar wind nebulae, seen across the electromagnetic spectrum in synchrotron and inverse Compton emission and in optical emission lines when they shock the surrounding medium. PSR¡¯s relativistic particle winds make SNR brighter. So the PWN can be seen as a strong evidence of interaction between a PSR and a SNR which also becomes a reliable sign of  association.\\
\\
This work is funded in part by ``the Fundamental Research Funds for the Central Universities" and by Graduate School of Beijing Normal University.
We thank supports from NSFC (011241001,11178007) and Bairen-program of the CAS.
\begin{table}
\tiny{ \begin{flushleft}
\centering
\caption{Candidates for PSR-SNR Associations\cite{tabel1}}
\begin{tabular}{cccccccccc}
  \hline\hline\hline
SNR Name               & Dis.(method) & Age & PWN and PSR                 & Dist.    & Age(real)  \\
                       & /kpc         & kyr &                             & kpc      & kyr         \\
  \hline\hline\hline
G5.27-0.9     & 4.3(f)         & ?            & PWN B1757-24                 & 5.8-1.3      & 15.5(39-170)\\
G7.5-1.7                 & $>$1.7(b)      & ?            & PWN $\gamma$-ray J1809-2332  & ?            & 67.6   \\
G8.7-0.1            & 4.5-6(e)       & $\sim$16     & J1803-2137                   & 3.9-5.3      & 15.8 \\
G11.2-0.3                & $\sim$5(f)     & $\sim$2      & PWN J1811-1925               & ?            & 24($\sim$2)\\
G12.82-0.2         & 4.7(b)         & $<$2.5       & PWN X-ray J1813-1749         & 4.7          & 3.3-7.5\\
G27.4-0.0         & 7.5-9.8(i)     & $\sim$2      & AXP 1E 1841-045              & 7.5          & 4.7($<$10)\\
G29.7-0.3         & 10.6(h)        & $\sim$1      & PWN J1846-0258               & ?            & 0.725\\
G34.7-0.4     & 2.5-2.6(e,f)   & $<$10        & PWN $\gamma$-ray? B1853+01   & 3.3          & 20.3\\
G40.5-0.5                & $\sim$3.4(h)   & 20-40        & PWN $\gamma$-ray J1907+0602  & 3.2          & 19.5\\
G54.1$+$0.3              & 4.5-9(f,h)     & 1.5-6.0      & PWN J930$+$1852              & $\sim$12     & 2.9\\
G57.3$+$1.2              & 5.4$\pm$1.9(a) & 40-290       & B1930$+$22                   & 6.6$\pm$2.0  & 39.8\\
G69.0$+$2.7      & 2(f)           & 77           & PWN $\gamma$-ray B1951+32    & 2.4$\pm$0.2  & 107(64$\pm$18)\\
G106.3$+$2.7             & 0.8?(f,h)      & ?            & PWN J2229$+$6114             & 0.8?         & 10.5\\
G109.1-1.0      & 4$\pm$0.8(f,h) & $\sim$10     & AXP 1E 2259$+$586            &4.6-5.1       &100-200\\
G114.3$+$0.3             &1.6-3.4(f)      & a few 10     &PWN B2334$+$61                &$\sim$2.5     &40.9\\
G119.5$+$10.2     &1.4$\pm$0.3(f)  &5-15          &PWN $\gamma$-ray 3EG J0010+7309&?            &$\sim$14\\
G130.7$+$3.1       &3.2(f)          &0.82          &PWN J0205$+$6449              &3.2           &5.4\\
N49                      &50              &5-16          &AXP and SGR J0526-66          &50            &2($\sim$6.6)\\
G180.0-1.7         &0.8-1.6(a,k)    &80-200        &PWN J0538$+$2817              &1.2           &600\\
G184.6-5.8     &2(p)            &0.94          &PWN $\gamma$-ray B0531+21     &2             &1.3\\
SNR 0538-691      &51              &$\sim$5       &PWN X-ray J0537-6910          &51?           &$\sim$5\\
SNR 0540-693      &55              &0.8-1.1       &PWN B0540-69                  &49.4          &1.67\\
G263.9-3.3     &0.22-0.28(m)    &5-29          &PWN $\gamma$-ray B0833-45     &0.24-0.37     &11.2(10)\\
G292.0$+$1.8  &3.6-5.5(k)      &2.40-2.85     &PWN J1124-5916                &5-6.8         &2.9\\
G310.6-1.6               &$\sim$7(b,k)    &$\leq$1       &PWN X-ray J1400-6326          &10$\pm$3      &12.7\\
G320.4-1.2     &3.8-6.6(f)      &6-20          &PWN $\gamma$-ray B1509-58     &4.4           &1.7\\
G328.4$+$0.2 &$\leq$17.4¡À0.9(f)&$\sim$7     &PWN neutron star?             &?             &7\\
G330.2$+$1.0             &$\sim$5-10(f)   &$\geq$3       &CXOU J160103.1.513353?        &$\sim$5-10    &$\leq$3\\
G332.4-0.4      &3.3-4.7(a,f)    &1-2           &J1617-5055                    &3.3           &1-8.1 \\
G341.2$+$0.9             &8.3-9.7(a)      &?             &PWN B1643-43                  &6.9           &32.6\\
G343.1-2.3    &$\sim$3(a)      &5-6           &PWN $\gamma$-ray B1706-44     &1.8-3.6       &17.5\\
G0.9$+$0.1               &$\sim$8.5(b)    &1-7           &PWN J1747.2809                &$\sim$13      &5.3($<$2.7)\\
\hline
G6.4-0.1            &2.5$\pm$0.7(d,f)&35-150        &J1801-23                      &9.4¡À2.4      &$<$58.3\\
G23.3-0.3           &3.9-4.5(i)      &$<$50         &B1830-08                      &4-5           &148\\
G29.6$+$0.1              &$<$20(d)        &$<$8          &AXP J1845-0258                &8.5-15        &?($\sim$10)\\
G33.6$+$0.1       &$\sim$7.8(f)    &5.4-7.5       &AXP J1852$+$0040              &7.1           &$>$24\\
G55.0$+$0.3 &14(f)           &$<$1500-2300  &J1932$+$2020                  &9.14          &1100\\
G65.1$+$0.6              &9.2(f)          &40-140        &J1957$+$2831                  &7.0$\pm$2.3   &1600(180)\\
G117.7$+$0.6             &3(f)            &10-20         &RX J0002$+$6246               &3.5           &9.12\\
G132.7$+$1.3        &2-3(a,d)        &21            &J0215$+$6218                  &2.3-3.2       &13100\\
G160.9$+$2.6        &1.1-4.0(i)      &a few 100     &B0458$+$46                    &1.8           &1813\\
G203.0$+$12.0            &0.1-1.3(l)      &86            &B0656$+$14                    &0.2-0.8       &110.8\\
G266.2-1.2    &0.2-1.0(n)    &0.7-1         &J0855-4644                    &0.25-0.75     & 140\\
G290.1-0.8   &7-8(h)          &10-20         &J1105-6107                    &7             &63 \\
G292.2-0.5               &8.4(f)          &2.9?          &J1119-6127                    &2.4-8         &1.6\\
G296.5$+$10.0 &1.3-3.9(d,f)&3-21          &RQNS 1E 1207.4-5209          &2             &340$\pm$140\\
G296.8-0.3      &9.6$\pm$0.6(f)  &2-10          &J1157-6224                    &$\sim$10      &1600\\
G308.8-0.1               &4.0-6.5(e)      &$<$32.5       &PWN? J1341-6220               &8             &12\\
G312.4-0.4               &2 or 3(a)       &?             &J1413-6141                    &2             &14\\
G327.24-0.13?            &3.7-4.3(b)      &?             &AXP1E1547.0-5408              &4-8           &1.41\\
G332.4$+$0.1      &7.5-11(b)       &$\sim$5       &B1610-50                      &7.24          &7.45\\
G335.2$+$0.1             &6(a)            &?             &J1627-4845                    &5.1-8.5       &2700\\
G347.3-0.5               &$\sim$1(h)/$\sim$6(d,h)&$\sim$10&J1713-3949                   &5.0$\pm$0.2   &$\sim$100\\
G354.1$+$0.1             &4.7-5.6(d)      &?             &B1727-33                      &4.2           & 26\\
G359.23-0.82?            &$<$5.5(f)       &?             &J1747-2958                    &$\sim$2       &25.5\\
G10.0-0.3        &14.5(d,e)       &?             &SGR 1806-20                   &14.5$\pm$1.4  &?\\
G337.0-0.1      &11(d,e,f,h)     &$\leq$20      &SGR 1627-41                   &11            &? \\
\hline
G16.8-1.1                &6.7(a)          &?             &B1822-14                      &$\sim$3.5     &200\\
G21.5-0.9                &$\sim$4.8(i)    &0.2-1         &J1833-1034                    &3.3-3.7       &4.9\\
G24.7$+$0.6              &4.4(a)          &12            &B1832-06                      &6.3           &120\\
G32.4$+$0.1              &$\sim$17(o)     &?             &J1850-0006                    &7.21          &8040\\
G42.8$+$0.6              &10$\pm$3(a)     &a few 10      &J1907-0918                    &7.7           &38\\
G78.2$+$2.1   &1.5$\pm$0.5(f)  &6.6           &$\gamma$-ray J2021$+$4026     &?             & $\sim$77\\
G82.2$+$5.3        &1.3-1.9(a)      &?             &W63 X-1                       &?             &?\\
G189.1$+$3.0&0.7-2(b,j)      &30            &PWN? B0611$+$22               &4.7           &$\sim$80\\
G260.4-3.4    &2.2$\pm$0.3(f)  &3.7$\pm$0.4   &J0821-4300                    &2.2           &$>$220\\
G284.3-1.8    &1-2.9(j,h)      &$\sim$10      &J1016-5857                    &7-12          &21 or 16\\
G309.8-2.6?              &?               &0.7-2         &J1357-6429                    &4             &7.3\\
G313.4$+$0.2?            &?               &?             &J1420-6048                    &8             &13\\
G336.1-0.2               &?               &?             &J1632-4818                    &8             &20\\
G348.7$+$0.3   &$\sim$8(f)      &$\sim$1.5     &CXOU J171405.7.381031         &?             &?\\
G350.1-0.3               &4.5(i)          &$\sim$0.9     &XMMU J172054.5-372652         &?             &?\\
G352.2-0.1               &?               &?             &J1726-3530                    &10            &14\\
G353.6-0.7               &3.2(b)          &$\sim$27      &XMMU J173203.3.344518         &?             &?\\
G354.8-0.8               &8(a)            &?             &J1734-3333                    &7             &8.1\\
G18.0-0.7?               &?               &?             &J1826-1334                    &3.4-4.3       &21.4\\
G111.7-2.1 &3.3-3.7(l)      &0.34          &CXO J2323$+$5848              &?             &?\\
G346.5-0.1?              &5.5 or 11.0(e)  &?             &AXP J1708-4009                &?             &?($<$10)\\
\hline\hline\hline
\\
\end{tabular}

\end{flushleft}}
\end{table}
\end{document}